\documentclass[11pt,preprint]{aastex}
\usepackage{amsmath}    
\usepackage{mathtools}
\usepackage[english]{babel}
\usepackage{graphicx}
\renewcommand{\vec}[1]{\mathbf{#1}}

\slugcomment{draft version \hskip 1.5cm \today}

\shorttitle{Collisionless Weibel shocks}
\shortauthors{Ardaneh et al.}

\begin{document}


\title{Collisionless Weibel shocks and electron acceleration in gamma-ray bursts}


\author{Kazem Ardaneh and Dongsheng Cai}
\affil{ CAVE Lab, Department of Computer Science, University of Tsukuba, Ibaraki 305-8573, Japan}
\email{kazem.arrdaneh@gmail.com}

\author{Ken-Ichi Nishikawa}
\affil{Department of Physics, University of Alabama in Huntsville, ZP12, Huntsville, AL 35805, USA}

\and

\author{Bertrand Lemb\'{e}ge}
\affil{LATMOS, Universit\'{e} de Versailles Saint-Quentin-en-Yvelines (UVSQ), 11 Boulevard D'Alembert, 78280 Guyancourt, France}

\begin{abstract}
A study of collisionless external shocks in gamma-ray bursts is presented. The shock structure, electromagnetic fields, and process of electron acceleration are assessed by performing a self-consistent 3D particle-in-cell (PIC) simulation. In accordance with hydrodynamic shock systems, the shock consists of a reverse shock (RS) and forward shock (FS) separated by a contact discontinuity (CD). The development and structure are controlled by the ion Weibel instability. The ion filaments are sources of strong transverse electromagnetic fields at both sides of the double shock structure over a length of 30 - 100 ion skin depths. Electrons are heated up to a maximum energy $\epsilon_{\rm ele}\approx \sqrt{\epsilon_{\rm b}}$,  where $\epsilon$ is the energy normalized to the total incoming energy. Jet electrons are trapped in the RS transition region due to the presence of an ambipolar electric field and reflection by the strong transverse magnetic fields in the shocked region. In a process similar to shock surfing acceleration (SSA) for ions, electrons experience drift motion and acceleration by ion filament transverse electric fields in the plane perpendicular to the shock propagation direction. Ultimately accelerated jet electrons are convected back into the upstream. 
\end{abstract}


\section{Introduction}
Relativistic jets are present in a variety of high-energy astrophysical systems, e.g., pulsar wind nebulae (PWNe), gamma-ray bursts (GRBs), and active galactic nuclei (AGNs). The jets propagate through an external medium and create a double shock structure at the leading edge \citep{bro11}. The most extreme relativistic jets are known to be associated with GRBs \citep{ram02}. GRBs are often seen when massive stars culminate their lives producing powerful internal jets and supernovae explosions. Non-thermal emission associated with the relativistic jets is believed to be generated by the accelerated electrons via synchrotron or inverse Compton mechanisms. Diffusive shock acceleration (DSA) is the most applicable theory for the acceleration of particles in shock systems \citep{blan87}. In this process, particle acceleration takes place as the result of particle trapping and oscillating at the shock front. Even though DSA is successful for elucidating  ion acceleration, acceleration of the electrons cannot be directly related to this mechanism. To experience DSA a thermal electron requires crossing the shock front multiple times. However, a thermal electron is closely tied to the magnetic field lines because of its small Larmor radius. Thus with no strong pre-acceleration process, a thermal electron is convected downstream without feeling any serious DSA. The lack of DSA for thermal electrons is often referred to as the electron injection problem \citep{hos02,ama09,guo14}.
 
PIC simulations can shed light on the mechanism of electron acceleration in collisionless shocks. In the high Mach number (the ratio of the upstream bulk flow speed to the sound speed in the ambient medium) regime, many authors have reported that the SSA process can inject high-energy electrons into the DSA \citep{die20,hos01,mcc01,hos02,sch02,ama09,mat12}. In the SSA process predominantly electrostatic fluctuations are excited at the leading edge of the shock via the Buneman instability \citep{bun58}, due to the interaction between incoming electrons and reflected ions. The incoming electrons are then trapped in the electrostatic potential and efficiently accelerated by the shock convective electric field (i.e., the electric field resulting from the motion of the magnetized upstream region toward the shock, $\vec{E}_0= -\vec{\beta_0} \times \vec{B_0}$).

To date, electron acceleration in unmagnetized ($\vec{B}_0=0$) electron-ion shocks has been poorly investigated. The injection process is anticipated to be different from magnetized shocks because there is no convective electric field to accelerate the electrons. In the present work, a 3D PIC simulation is used to study the shock structure, and the process of electron acceleration associated with a relativistic jet propagating into an unmagnetized plasma. This setup is the most applicable model for external GRBs shocks. The spatial and temporal scales of the present simulation are much larger than those used in previous studies \citep{nish03,ard14,cho14} in order to allow us to observe efficient particle acceleration. The present paper aims to address the following questions related to an unmagnetized double shock system: (i) What are the field structures responsible for the processes of electron heating, and acceleration?, (ii) Where do these processes mainly take place?, (iii) What is the resulting associated electron energy spectrum?, and (iv) What are the principal mechanisms responsible for electron heating, and acceleration?

In this paper the simulation method and parameters setup are summarized in $\S$ 2. The main simulation results are discussed in $\S$ 3. Finally, we conclude in $\S$ 4.

\section{Simulation method and parameter setup}

In recent works, in order to generate a shock, a relativistic plasma stream is injected from one end of the computational grid and reflected from a rigid wall at the opposite end (for instance, a 1D simulation by \citet{hos02}, 2D simulations by \citet{spi8a,spi8b, ama09,mar09}, and a 3D simulation by \citet{guo14}). This method resembles the collision of two identical counter-streaming beams and reduces by one-half the number of calculations. This approach simulates only a moving FS. In the present work we inject a particle jet into an ambient plasma, so that a double shock structure is fully captured. With this setup the jet-to-ambient density ratio can be changed and by assuming a density ratio greater than one, the shock formation process can be properly handled by smaller scale (temporal and spatial) simulations. Deceleration of the jet flow by the ambient plasma results in a CD (the CD is the location where the electromagnetic field and the velocity of jet and ambient plasmas are similar and the density changes). Two shocks propagate away from the CD into the jet and ambient upstreams (in the CD frame). Hence, two shocks and one CD split up the jet and ambient plasma into four regions: (1) unshocked ambient, (2) shocked ambient, (3) shocked jet, and (4) unshocked jet. Hereafter, subscripts 1, 2, 3, and 4 refer to the unshocked ambient, shocked ambient, shocked jet, and unshocked jet, respectively. Quantities with a single index $Q_{\rm i}$ denote the value of quantities $Q$ in region $i$ in frame $i$ and quantities with double indices $Q_{\rm ij}$ denote the value of quantities $Q$ in region $i$ as seen in frame $j$. 

The present simulation is performed using an adopted version of the TRISTAN code \citep{bun93,niem08}, a massively parallel, fully relativistic, PIC code that can be used for many applications in astrophysical plasmas \citep{nish03,nish05,nish09,ard14,cho14}. The simulation is carried out in the ambient frame with a box including 8005 nodes in the $x$-direction, and 245 nodes in the $y$ and $z$-directions. There are 6 particles per cell per species for the ambient plasma ($\approx$ 3 billion particles per species) and the jet-to-ambient density ratio used is 1.7. The ion-to-electron mass ratio used is 16. The system is numerically resolved with $5\Delta x$ per electron skin depth in each direction and $\Delta t\omega_{\rm pe}=0.01$, where $\Delta x$ and $\Delta t$ are the grid size and the time step, respectively. The thermal speed of ambient electrons and ions is $0.05c$ and $0.01c$, respectively. In the simulation, the jet fills the whole computational box in the $yz$-plane and is injected at $x = 25\Delta x$ in the positive $x$-direction.  The injected jet bulk speed is $\beta_{41}=0.9950$, $\gamma_{41}=(1-\beta_{41}^2)^{-1/2}=10$, the Mach number $M\equiv\beta_{41}/\beta_{\rm s1}=60$, and the jet electrons and ions have a thermal speed of $0.014c$ and $0.003c $, respectively. The surfaces at $x = x_{\min}$ and $x_{\max}$ are rigid reflecting boundaries for the ambient particles, while they are open boundaries for the jet particles. These surfaces are radiating boundaries for the fields based on Lindman's method as explained in \citet{bun93}. Periodic boundary conditions are applied at all other boundaries for both particles and fields. In the following, time is normalized to $\omega_{\rm pe}^{-1}$, space to the electron skin depth, $\lambda_{\rm ce}=c/\omega_{\rm pe}$, particle momentum for species $i$ to the corresponding $m_{\rm i}c$ (e: electron and i: ion), and density to the upstream ambient density, $n_1$. The dimensionless form of a variable, such as $x$, is shown by $x^*$. The energy is normalized to the total incoming energy density, $\sum{n_{\rm n}\gamma_{\rm n}m_{\rm n}c^2}$, and is denoted by $\epsilon$, according to the notation of \citet{med99}.

\section{Simulation Results}

The present simulation includes growth of the Weibel instability, and the generation of electromagnetic fields which decelerate the jet flow and consequently form a double shock structure. At late times the particles are effectively heated, and accelerated. In the following this scenario is clarified in more detail.

\subsection{Shock structure}

\begin{figure}[h]
\begin{center}
\includegraphics[scale=0.5]{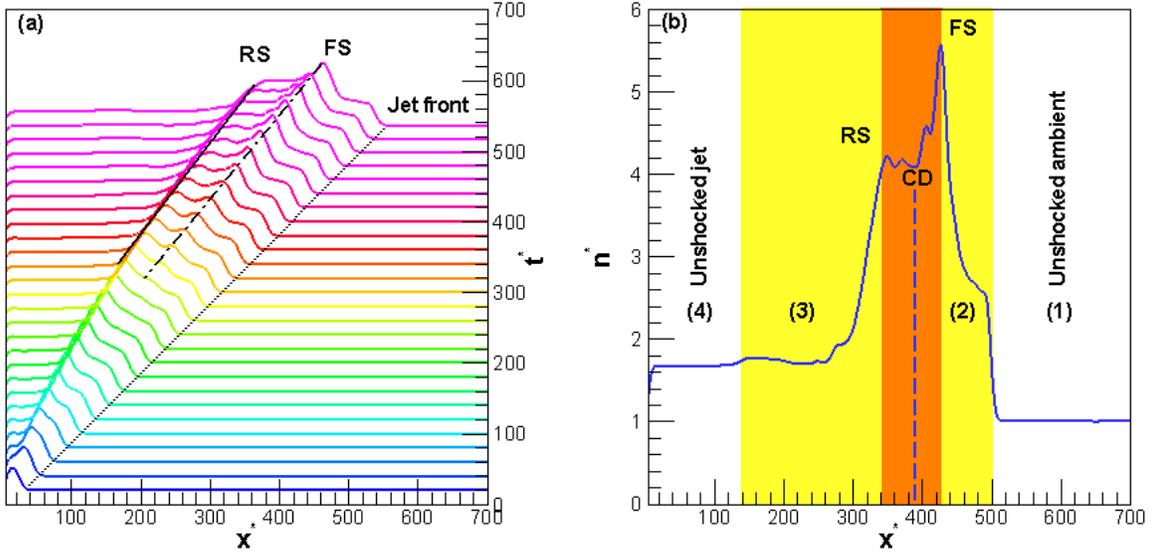}
\caption{Panel (a) shows stacked profiles of the transversely averaged ion density (total jet+ambient) from 
$t^*=20$ up to $540$ with an interval of $\Delta t^*=20$ and panel (b) represents the ion density profile at $t^*=500$. Solid, dashed-dot, and dotted black lines in panel (a) indicate the RS, FS, and jet front, respectively. In panel (b) the shocked region is shown as orange, yellow represents the shock transition region (left for the RS and right for the FS), white shows the unshocked regions (jet or ambient), and the dashed blue line shows the CD.}
\label{sh}
\end{center}
\end{figure}

When the jet with density ratio of $n_{41}/n_1=1.7$ and $\gamma_{41}=10$ interacts with the ambient plasma, the total density initially increases to $2.7n_1$ near the jet front. In this interaction, the distribution of particles is extremely anisotropic and this distribution is unstable to several plasma instabilities, such as the electrostatic Buneman instability \citep{bun58} and the electromagnetic Weibel instability \citep{wei59}. In relativistic shocks, the Weibel instability has the largest growth rate and will dominate the interaction \citep{med99,cal02,hed05}. Time evolution of the Weibel instability was analyzed in \citet{ard14} and we summarize the results here. Jet electrons are deflected when interacting with ambient particles to form electron current filaments. Subsequently magnetic fields are amplified principally as a result of mutual attraction between the electron filaments and their merging to larger scale filaments. When the magnetic fields become strong enough to deflect the heavier ions, the ions begin to participate in the instability. The ion current filaments then grow with a smaller growth rate related to the electron-ion mass ratio and to transverse heating. As time evolves, the Weibel instability induced electromagnetic fields facilitate momentum transfer between the jet and ambient plasma. A double shock structure forms when the relative velocity between the jet and ambient plasmas exceeds the sound speed in the ambient plasma and the magneto-sonic speed in the jet, and the pressure in the shocked region exceeds the pressure in the unshocked region \citep{zha05}. A first density compression appears due to the deflection of jet electrons and ambient ions which we define as the RS. This density compression grows due to amplification of the Weibel instability at the expense of the streaming ions. Time stacked plots of the transversely averaged (in the $yz$-plane) total ion density as a function of axial distance are shown in Fig. \ref{sh}a. The time range is $t^*=20-540$ with an interval of $\Delta t^*=20$. One can readily identify a RS propagating in the positive $x$-direction with $\beta_{\rm RS1}\approx0.70$. The shape and peak value of the total density corresponding to the RS are almost constant in time ($n_{31}/n_{41}\approx2.35$, Fig. \ref{sh}b), and the RS is fully developed. On the other hand, a density compression appears mainly in the ambient plasma at late times ($t^*\approx300$) that we identify as the FS. The compression ratio rises with time until reaches about $n_{21}/n_1\approx6$ at the end of the simulation (Fig. \ref{sh}b). The FS structure moves at a speed $\beta_{\rm FS1}\approx0.89$ in the positive $x$-direction. The CD moves in the positive $x$-direction at a speed $\beta_{\rm CD1}\approx0.80$. The shocked region ($340\leq x^*\leq430$ at $t^*=500$) between the RS and FS separates the jet and ambient upstreams (Fig. \ref{sh}b). 

The results of present simulation are quantitatively validated by the theoretical calculations described in the Appendix. The speed of the CD in the ambient frame from Eq. (\ref{scd})) is $\beta_{\rm CD1}=0.8$. Combining Eq. (\ref{sgb}a) and (\ref{sgb}b), the FS speed can be determined. For a shocked ambient adiabatic index between $4/3\leq\tilde{\Gamma}\leq 5/3$ and $\gamma_{\rm CD1}=1.67$ (corresponding to $\beta_{\rm CD1}=0.8$) the FS speed will lie between  $0.85\leq\beta_{\rm FS1}\leq0.90$ ($1.91\leq\gamma_{\rm FS1}\leq2.23$) where upper and lower limits correspond to the upper and lower limits of  $\tilde{\Gamma}$, respectively. From Eq. (\ref{fjc}), the jump condition at the FS lies between  $5.66n_1\leq n_2 \leq 9.66n_1$, where  $n_2$ is the shocked ambient density in the CD frame and the lower and upper limits correspond to the upper and lower limits of $\tilde{\Gamma}$, respectively. Measured in the ambient frame the density compression would lie between
$9.4n_1\leq n_{21}\leq16.0n_1$.  In the jet frame the CD moves with speed $\beta_{\rm CD4}=-\beta_{\rm 4CD}=-0.96$ (Eq. (\ref{cds}a)) and the corresponding Lorentz factor is $\gamma_{\rm CD4}=3.41$. By $ 1\rightarrow4$, $2\rightarrow3$ and $\gamma_{34}\rightarrow\gamma_{\rm CD4}$ transformations in Eqs. (\ref{sgb}), we can calculate the speed and the corresponding Lorentz factor of the RS in the jet frame. In this frame we have $ 0.97\leq-\beta_{\rm RS4}\leq0.98$ and $4.35\leq\gamma_{\rm RS4}\leq5.77$, where upper and lower limits correspond to upper $\tilde{\Gamma}=5/3$ and lower   
$\tilde{\Gamma}=4/3$ limits for the shocked jet adiabatic index, respectively. The RS speed in the ambient frame (Eq. (\ref{rss})) lies between $0.51\leq \beta_{\rm RS1} \leq 0.70 $. The density jump at the RS region can be obtained from Eq. (\ref{rjc}) where $n_4=n_{41}/\gamma_{41}$. Hence the density compression of shocked jet is between $1.0n_{41}\leq n_3\leq 1.66 n_{41}$ according to the upper and lower limits of $\tilde{\Gamma}$. In the ambient frame the shocked jet density lies between $1.7 n_{41} \leq n_{31}\leq 2.7 n_{41}$. Although the density jump for the shocked ambient is about a factor of $\sim1.5$ smaller than theoretically predicted for a fully developed FS, the simulation results for the RS, CD, and FS speed and also for the density jump at the RS are in the ranges provided by the theoretical analysis for a fully developed shock system. 
 
\begin{figure}[h]
\begin{center}
\includegraphics[scale=0.33]{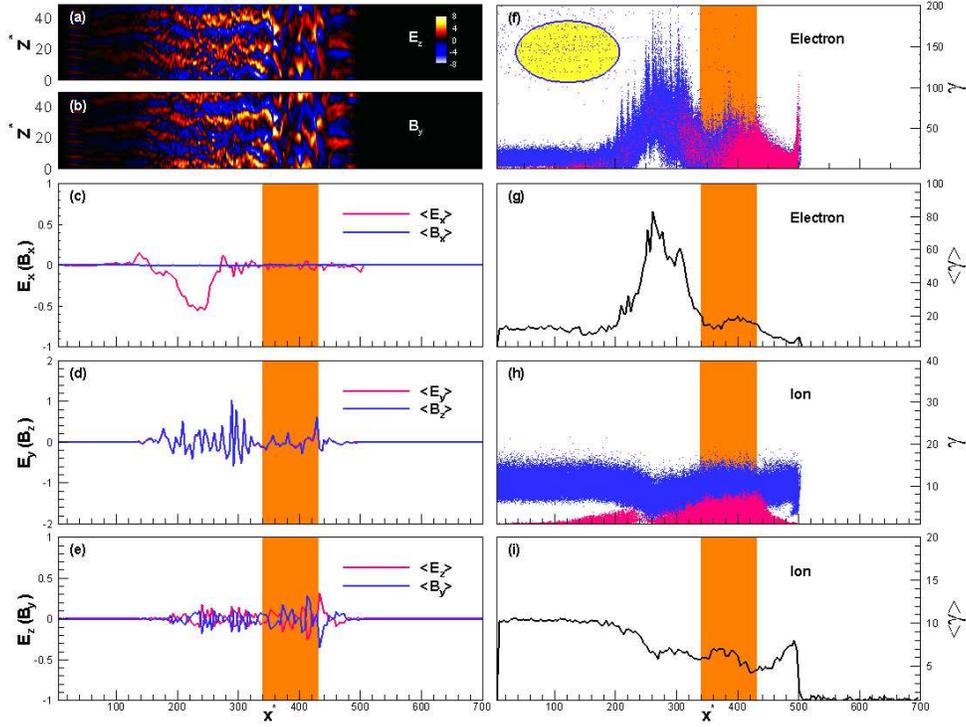}
\caption{Structure of the relativistic double shock at $t^*=500$. Panels (a) and (b) show the 
$z$-component of the electric field $E_{\rm z}$ and the $y$-component of the magnetic field $B_{\rm y}$ at $y^*=24$. Panels (c)-(e) show the transversally averaged (in the $yz$-plane) field components. Total of the jet (blue) + ambient (red) particle energy distribution, and average energy along $x$-direction are presented in panels (f) and (g) for electrons, and (h) and (i) for ions. The shocked region is identified by orange coloring or between vertical orange lines (as shown in Fig. \ref{sh}). The ellipse in panel (f) shows the high-energy electrons reflected into the upstream. In (f) and (h), $1.2\times10^6$ particles are randomly selected. Due to the very large number of particles in the simulation, only a part of them is represented. Particles of each population are selected randomly so that the respective distribution function is not affected.}
\label{prof}
\end{center}
\end{figure}

Shown in Fig. \ref{prof}a and \ref{prof}b are the $z$-component of the electric field, $E_{\rm z}$, and $y$-component of the magnetic field, $B_{\rm y}$, respectively. Panels \ref{prof}c-\ref{prof}e show the transversally averaged (in the $yz$-plane) electric and magnetic field components, $[E_{\rm x}(B_{\rm x}), E_{\rm y}(B_{\rm z}), E_{\rm z}(B_{\rm y})]$. The energy distribution (total of jet+ambient) and average energy along the $x$-direction for electron and ion species are shown in Figs. \ref{prof}f-\ref{prof}i. All panels are at $t^*=500$. Where high-speed jet particles interact with the ambient medium (behind the RS at $x^*\leq340$) or scattered ambient particles blend with the upstream ambient (in front of the FS at $430\leq x^*$), particles distribution becomes strongly anisotropic. Anisotropies result in the Weibel instability which generates current filaments in these regions with currents in the $x$-direction. According to Ampere's law, these current filaments are encircled by transverse magnetic fields, and we see that  $\langle B_{\rm x}\rangle=0$ in Fig. \ref{prof}c.  The transverse electric fields are related to the magnetic fields via $\vec\beta_{\rm e:i}\times\vec{E}=\vec{B}$ where $\beta_{\rm e:i}$ is the velocity of the electron (ion) carrier. The carriers move roughly at the speed of light in the $x$-direction, $\beta_{\rm e:i}\simeq\beta_{\rm e:ix}\simeq1$. Therefore, the transverse electric field components are $E_{\rm y}=B_{\rm z}$, and $E_{\rm z}=-B_{\rm y}$, as are observed in the simulation results for $[E_{\rm y}(B_{\rm z}), E_{\rm z}(B_{\rm y})]$ (Figs. \ref{prof}a-\ref{prof}e). Additionally, there is a longitudinal ambipolar electric field within the RS transition region, $140\leq x^*\leq340$ for $t^*=500$ (Fig. \ref{prof}c). This electric field is generated by the density gradient and different mobilities of electrons and ions \citep{for70,for71,hos01,cho14}. The magnetic fields act to isotropize the momentum distribution, while the electric fields function to thermalize, and accelerate the particles afterwards. In Figs. \ref{prof}f-\ref{prof}i, the shocked region lies between $x^*=340$ and $x^*=430$. Within the RS transition region ($140\leq x^*\leq340$) jet electrons are trapped by the ambipolar electric field and effectively accelerated up to $\gamma_{\rm e}=200$ by the transverse electric fields (Figs. \ref{prof}f and \ref{prof}g). A tenuous population of these electrons convect upstream due to reflection by the magnetic fields in the shocked region (ellipse in Fig. \ref{prof}f). On the other hand,  jet ions are slowed in the RS transition region ($140\leq x^*\leq340$) by 40\% from the initial Lorentz factor $\gamma_{\rm i}=10$, due to the effect of the ambipolar electric field. In the shocked region, jet electrons have been fully thermalized and are well merged with the thermalized ambient electrons. Thus, only a single electron population is present in the hot shocked region (Fig. \ref{prof}f). On the other hand, the kinetic energy of jet ions is transferred to the heating of ambient particles by means of the electromagnetic fields generated by the ion Weibel instability (Fig. \ref{prof}h). Full thermalization of the two ion populations (jet and ambient) has not yet occured (demands a longer simulation time), i.e., the two populations are distinguishable in Fig. \ref{prof}h. Electrons located in the FS transition region ($430\leq x^*\leq500$), predominantly ambient electrons, also undergo the Weibel instability and are thermalized by the jet upstream kinetic energy. In this region, penetrating jet ions interact with ambient particles and are slowed down gradually by 50\% from the jet front Lorentz factor $\gamma_{\rm i}=10$ to a minimum value of $\gamma_{\rm i}=5$ (Figs. \ref{prof}h and \ref{prof}i).

\begin{figure}[h]
\begin{center}
\includegraphics[scale=0.4]{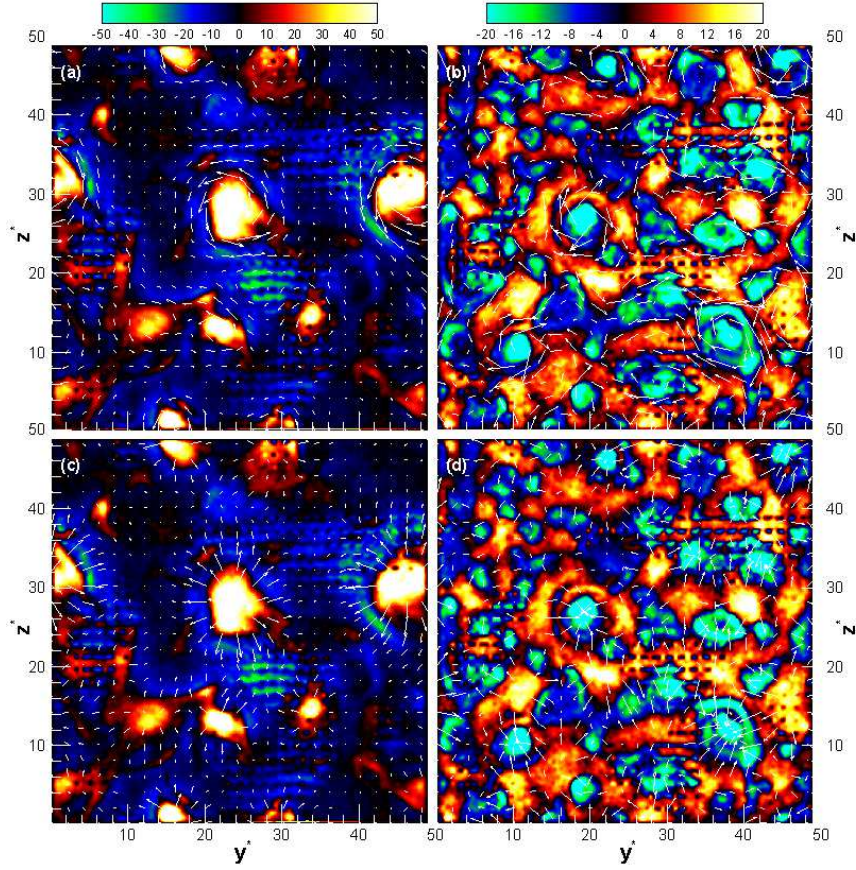}
\caption{Longitudinal current density through a transverse cross section at $x^*=320$ (first column) and $x^*=480$ (second column) for $t^*=500$. The arrows represent the transverse magnetic (first row), and electric (second row) fields.}
\label{cur}
\end{center}
\end{figure}

Ion current filaments generated by the ion Weibel instability play a crucial role in establishing the shock transition, electron heating, and acceleration in an unmagnetized electron-ion plasma \citep{fre04,hed04,spi8a}. In Figs. \ref{cur}, the longitudinal current density, $J_{\rm x}=J_{\rm ix}+J_{\rm ex}$, through a transverse cross section is shown at $x^*=320$, in the RS transition region (Fig. \ref{cur}a and \ref{cur}c) and at $x^*=480$, in the FS transition region (Fig. \ref{cur}b and \ref{cur}d) for $t^*=500$. Positive (red-white colors) and negative (green-blue colors) represent the ion and electron contribution to the total current, respectively. As one can see, longitudinal current filaments are surrounded by approximately azimuthal magnetic fields (illustrated by the arrows in Fig. \ref{cur}a and \ref{cur}b). Electric fields are perpendicular to the magnetic fields and the associated arrows point in the radial direction (Fig. \ref{cur}c and \ref{cur}d). In the RS transition region, magnetic fields are predominantly due to the ion filaments (counter-clockwise arrows) and electrons act to Debye shield these filaments \citep{fre04}. However, there are some electron filaments (clockwise arrows) within the FS transition region. Furthermore, electromagnetic fields in the RS transition region are stronger than those in the FS transition region. The transverse size of the filamentary structures is on the order of several electron Larmor radii ($R^{*}\approx3-5$). Perpendicular electric and magnetic fields lead to the $\vec{E} \times\vec{B}$ drift motion of electrons parallel to the shock propagation direction ($x$-direction). During this motion, the electrons are effectively heated.

\subsection{Acceleration mechanism}

\begin{figure}[h]
\begin{center}
\includegraphics[scale=0.5]{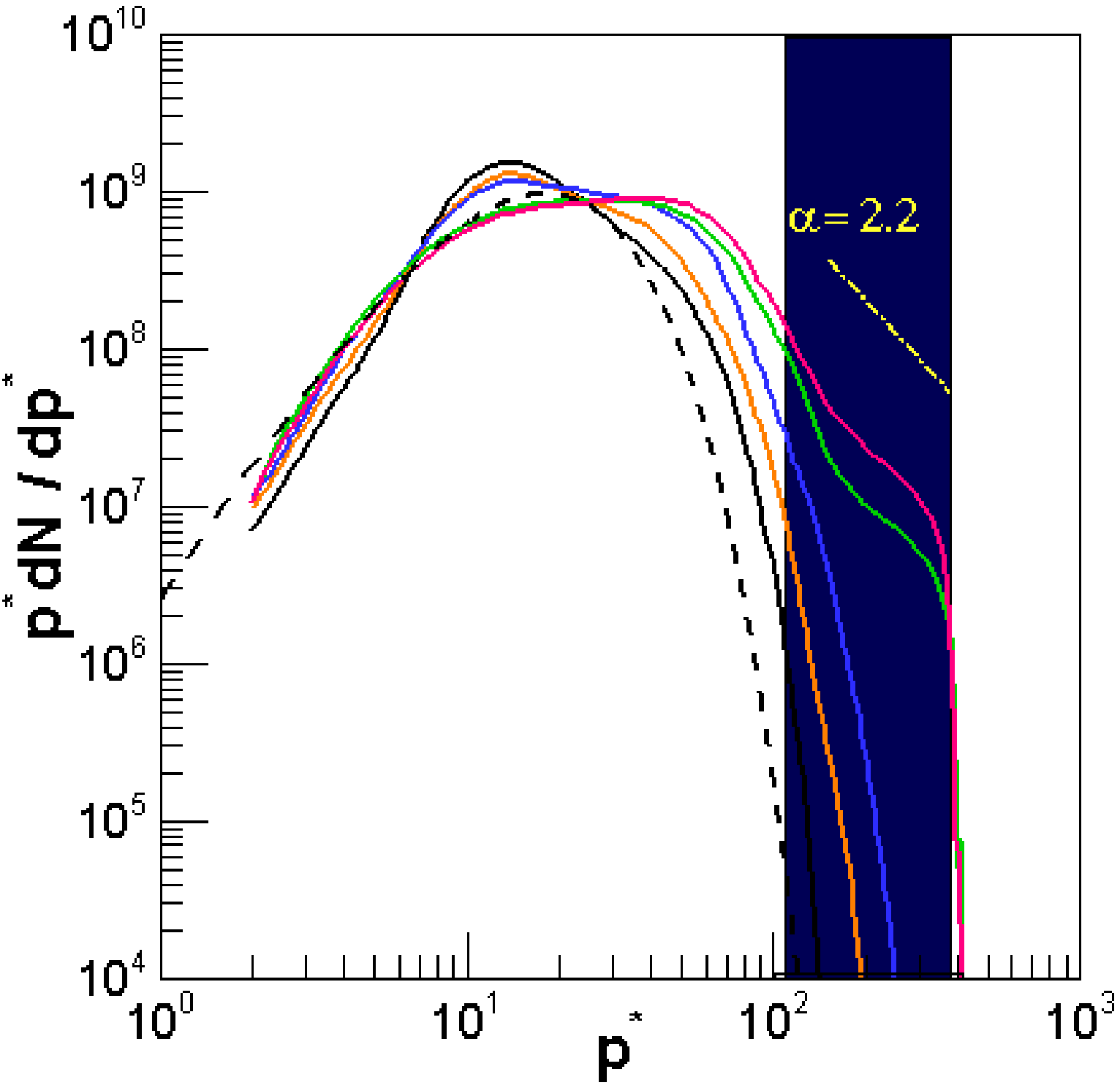}
\caption{The electron spectrum taken at $200\lambda_{\rm ce}$ within the RS transition region from $t^*=420$ (leftmost solid black line) up to 500 (rightmost solid red line) with an interval of $\Delta t^*=20$ ($200\lambda_{\rm ce}$ is calculated relative to the RS position at each time). The dashed line shows a 3D Maxwell-J\"{u}ttner distribution. Dashed-dotted line shows a power-law fit to the non-thermal component in the electron energy spectrum at the latest time.}
\label{spectrum}
\end{center}
\end{figure}

One of the key factors in generating non-thermal radiation from GRB afterglows is a non-thermal, high-energy electron population. This population can be seen in the electron spectrum, where a pure 3D Maxwell-J\"{u}ttner distribution does not account for the high energies. In fact, a more complex spectrum is anticipated as a result of electron acceleration. Shown in Fig. \ref{spectrum} is the electron spectrum from time $t^*=420$ up to 500 with an interval of $\Delta t^*=20$ taken at $200\lambda_{\rm ce}$ in the RS transition region where most of the accelerated electrons are located. The electron spectrum consists of a main thermal peak around $p^*\approx14$, and a high-energy tail which extends to $p^*\approx400$ at $t^*=500$. The dashed line shows a 3D Maxwell-J\"{u}ttner distribution with $f(p^*)={p^{*2}}/{T^*K_2(1/T^*)}\exp{(-{\sqrt{1+p^{*2}}}/{T^*})}$ where $K_2$ is the modified Bessel function of the second kind, and $T^*=K_{\rm B}T/m_{\rm e}c^2$. We set $T^*=5$, in accordance with $T^*=(\gamma_{41}-1)n_{41}/3n_1=5$ which is expected if the upstream jet kinetic energy is completely converted into internal kinetic energy of the jet and ambient particles. The dashed-dotted line shows a power-law fit to the non-thermal, high-energy electron population. The power-law begins around $p^*=100$ and extends to $p^*=360$ with index $\alpha=2.2$, and is in agreement with theoretical predictions for a $\gamma =10$ relativistic shock \citep{kir20,ach01}. 

\begin{figure}[h]
\begin{center}
\includegraphics[scale=0.5]{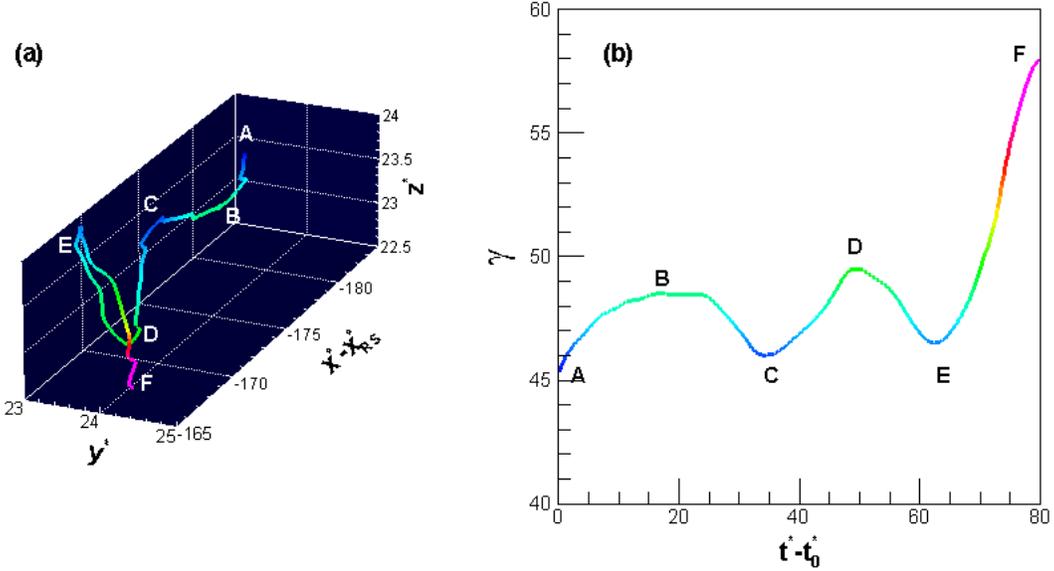}
\caption{Panel (a) shows the 3D trajectory of an electron (selected at $t_0^*=420$) during heating over a time interval $\Delta t^*=80$ and panel (b) shows the electron energy over time. The color of the electron trajectory expresses its energy based on the color in panel (b). The longitudinal electron position is calculated relative to the RS front position $(x_{\rm e}^*-\beta_{\rm RS1}\times t^*)$.}
\label{heating}
\end{center}
\end{figure}

The mechanism responsible for the non-thermal tail is analyzed by tracing the trajectories of some of the electrons located in the RS transition region at $t^*=420$. At first, the electrons are heated with heating principally associated with the ion current filaments \citep{hed04,med06,spi8a}. As discussed previously, when a relativistic jet interacts with the ambient medium, current filaments are formed by the Weibel instability. The ion filaments are sources of strong electromagnetic fields in the transition regions of the RS and FS, while the electrons provide Debye shielding of the ion filaments. The electromagnetic fields in the vicinity of the ion current filaments can be seen in Figs. \ref{cur}. The magnetic fields are predominantly azimuthal while the electrostatic fields are in the radial direction. The presence of perpendicular electric and magnetic fields results in a drift motion of the electrons with $v_{\rm d}=|\vec{E}\times\vec{B}| /B^2$ parallel to the shock propagation direction ($x$-direction). At distances less than a Debye length, an electron initially moves in a direction opposite to the electric field. Due to the velocity $v_{\perp}$ thus acquired, the magnetic part of the Lorentz force produces a motion perpendicular to the electric and magnetic fields which instantly bends the electron trajectory. During the first-half gyration, an electron is accelerated at the expense of the potential energy stored in the transversal electric field (A-B in Fig. \ref{heating}a and Fig. \ref{heating}b). On the other hand, an electron in its second-half gyration does work on the electric field and its kinetic energy will be transferred to the electric field energy (B-C in Fig. \ref{heating}a and Fig. \ref{heating}b). The growth of the ion Weibel instability increases the electric charge of the ion filaments with time, hence in the next gyration (C-D-E in Fig. \ref{heating}a and Fig. \ref{heating}b) an electron experiences deeper electric potentials that enhance the amplitude of energy oscillations. The maximum attainable energy for an electron during the heating stage can be estimated analytically. Approximately, an electron moving toward an ion current filament gains energy $u_{\rm ele}\approx eE\Delta r \simeq eB\Delta r$ ($\beta_{\rm xi}\simeq1$). The maximum radial distance that an electron can travel is about half the distance between the filaments (measured from the filaments axes), $\Delta r\approx c/\omega_{\rm pi,r}$, where $\omega_{\rm pi,r}=(4\pi n_{\rm i}e^2/m_{\rm i}\gamma_{\rm i})^{1/2}$ is the relativistic ion skin depth. Hence, the electron energy density is $\epsilon_{\rm ele}\approx \sqrt{\epsilon_{\rm b}}$ (normalized to the total incoming energy). In the present simulation, the maximum of $\sqrt{\epsilon_{\rm b}}$ for $140\leq x^*\leq340$, where electrons are found at high energies, is about 0.06 (calculated based on the magnetic field components in Figs. \ref{prof}c-\ref{prof}e and normalized to the total incoming energy $\sum{n_{\rm n}\gamma_{\rm n}m_{\rm n}c^2}=170$). Therefore, the maximum Lorentz factor of electrons during the heating stage is about 102, in agreement with the 3D Maxwell-J\"{u}ttner distribution dashed line shown in Fig. \ref{spectrum}. 

\begin{figure}[h]
\begin{center}
\includegraphics[scale=0.35]{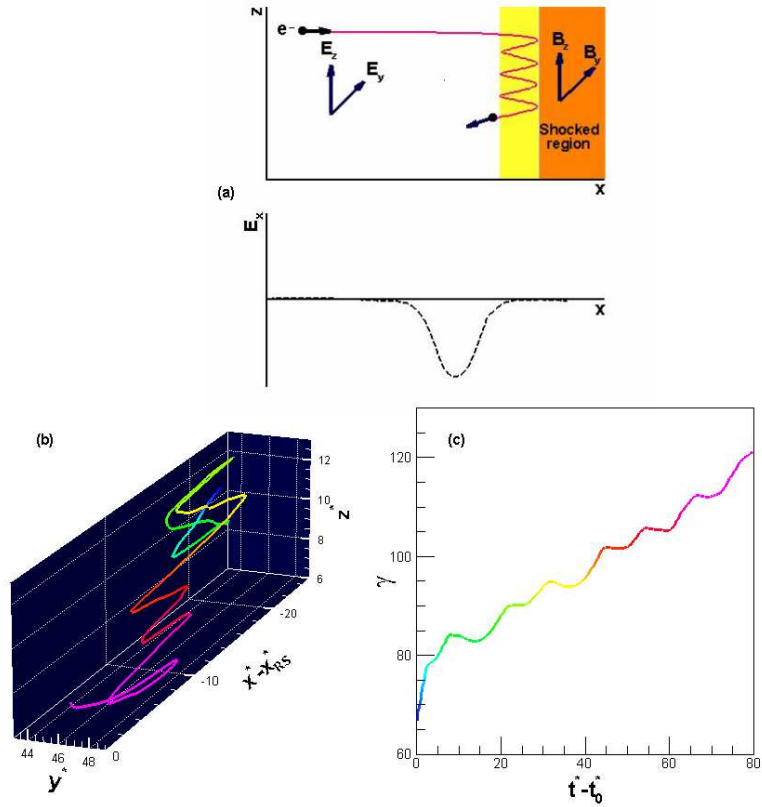}
\caption{Panel (a) represents the schematic diagram of the electron acceleration process, panel (b) is the 3D trajectory of an electron (selected at $t_0^*=420$) during acceleration in a time interval $\Delta t^*=80$, and in panel (c) the electron energy over time is illustrated. The colors of the electron trajectory reflect its energy according to the color in panel (c). The longitudinal electron position is calculated relative to the RS front position $(x_{\rm e}^*-\beta_{\rm RS1}\times t^*)$.}
\label{trajectory}
\end{center}
\end{figure}

Acceleration of electrons within the RS transition region (Fig. \ref{prof}f) is associated with an ambipolar large-scale electric field induced in the negative $x$-direction due to the inertia difference between electrons and ions \citep{hos01}. The basic idea behind the electron acceleration process is illustrated in Fig. \ref{trajectory}a. In the figure the yellow region represents a distance in front of the ambipolar electric field and behind the RS (such as $250\leq x^*\leq340$ in Fig. \ref{prof}f) in which electrons can oscillate. Consider an incoming electron that enters this region with a small $\beta_{\rm ex}$. The electron moves in the positive $x$-direction (opposite the ambipolar electric field direction), and will be reflected by the strong magnetic fields $(B_{\rm y}, B_{\rm z})$ in the shocked region, and then this process may repeat. During this process, the electron experiences a drift motion in the $yz$-plane parallel to the RS plane (Figs. \ref{trajectory}a and \ref{trajectory}b). As a result, the electron is efficiently accelerated by the transversal electric fields (Fig. \ref{trajectory}c) because for $|\vec{E}_{\perp}|=|\vec{B}_{\perp}|$ (Figs. \ref{prof}a-\ref{prof}e) and $0\leq\beta_{\rm ex}\leq1$, $| e\vec{E}_{\perp}| \geq | e\vec\beta_{\rm ex}\times\vec{B}_{\perp}|$. After multiple reflections, the magnetic force experienced by the electron, $e(\beta_{\rm ey}B_{\rm z}-\beta_{\rm ez}B_{\rm y})$ becomes larger than the electric force, $eE_{\rm x}$.  At this point the electron will be convected in the negative $x$-direction and again be trapped, but now in the region of the ambipolar electric field (like $140\leq x^*\leq250$ in Fig. \ref{prof}f). The electron undergoes the same acceleration process in this region until it is convected into the upstream. The resulting small high-energy electron population is indicated by the ellipse in Fig. \ref{prof}f. This acceleration process resembles the SSA process for ions \citep{sag74,zan96,lee96}. This efficient acceleration process for electrons in the shock transition region is indicated in 1D PIC simulations by \citet{hos01,hos02} and in 2D PIC simulations by \citet{ama09,mat12}), where a series of large amplitude electrostatic solitary waves (ESWs) were induced by the Buneman instability resulting from the interaction between reflected ions and incoming electrons. Instead of the upstream convective electric field in the SSA, in the present acceleration process there are transverse electric fields generated by the Weibel instability that effectively accelerate the electrons. Although previous work has shown that at late times both electrons and ions are accelerated by the DSA \citep{sir13}, in the present simulation, electrons are accelerated to the highest energies via the SSA process.

\section{Summary and conclusions}

In this work, we have studied the physics of electron acceleration in a relativistic collisionless Weibel instability mediated shock by means of a 3D PIC plasma simulation. In our simulation, an unmagnetized jet plasma with a 3D drifting Maxwell-J\"{u}ttner distribution and temperature $T_{\rm e} = T_{\rm i}$ is injected into an ambient unmagnetized medium. This process results in a double shock structure which consists of three regions. 1) A pre-shock acceleration region (behind the RS),  consisting of the incoming jet and ambient particles and with large particle distribution anisotropy. In this region, strong electromagnetic fields due to the Weibel instability, and an ambipolar electric field due to the density gradient and the different mobilities of electrons and ions are induced. 2) A fore-shock region (in front of the FS) where the magnetic fields are relatively weak but still strong enough to be considered for some GRB afterglow radiation. 3) A hot shocked region where the magnetic field energy density is a few percent of equipartition with the kinetic energy density. This region is clumpy, with large structures that are on the order of 30 ion skin depths in length. In the present simulation, the jet RS forms first. The ambient medium FS forms at later times. In this simulation, the RS is fully developed and the shock structure and jump conditions confirm theoretical predictions obtained from an analytical treatment. The FS still developing at the end of the simulation and has not yet reached the theoretically predicted jump conditions.

Ion current filaments are the origin of deeply penetrating electromagnetic field structures. Heating of electrons is directly tied to the ion filaments. This links electron heating closely with the micro-structure of relativistic collisionless shocks. With no restriction on the stage of shock formation, the maximum electron energy density is $\epsilon_{\rm ele}\approx \sqrt{\epsilon_{\rm b}}$ during the electron heating stage. This is in excellent agreement with values of $\epsilon_{\rm ele}/\sqrt{\epsilon_{\rm b}}$ derived from afterglow data from 10 GRBs \citep{pan01, med06}. Thus, there are two equipartition parameters in the GRB shock: the magnetic field energy density , $\epsilon_{\rm b}$, and the energy density of the electrons, $\epsilon_{\rm ele}$, normalized to the total energy density of the shock, respectively. In the current simulation, the electrons are heated by the Lorentz factor $\gamma_{\rm e}=100$.

We have shown that non-thermal electrons with a power-law index of 2.2 are generated behind the RS. Concerning a relativistic jet propagating into an unmagnetized plasma, this non-thermal electron population is observed for the first time to the knowledge of the authors. Although the particles are mostly accelerated in the shock-dominated systems via DSA, tracing the electron trajectories reveals that the mechanism responsible for electron acceleration differs significantly from DSA. Acceleration of electrons is due to the presence of a strong ambipolar electric field behind the RS. The incoming jet electrons move opposite to the direction of the ambipolar electric field ($x$-direction in the present work). The electrons are then reflected back by the strong transverse magnetic fields in the shocked region. This results in an electron drift motion in a plane parallel to the RS plane while they are efficiently accelerated by the transverse electric fields. When the magnetic force becomes large enough to overcome the force of the ambipolar electric field, the electron are convected into the jet upstream of the RS. Therefore, a high-energy tenuous population of electron exists in the jet upstream. The acceleration mechanism is similar to the SSA for ions, where transverse electric fields generated by the Weibel instability play the role of the upstream convective electric field in the SSA.

\acknowledgments

The authors thank the referees for insightful comments and suggestions that led to correct interpretation of the acceleration observed in the present simulation. The authors thank Philip Hardee for critical reading and improvement of the text. The work of KN is supported by NSF AST-0908040, NNX12AH06G, NNX13AP21G, and NNX13AP14G. The simulation presented here was performed using the supercomputer of ACCMS, Institute for Information Management and Communication, Kyoto University.

\appendix

\section{Jump conditions for a relativistic 90\arcdeg shock}

We review first the MHD analysis of a relativistic 90\arcdeg shock propagating into an unmagnetized plasma, and then extend the analysis to double shock structure. In the following, indices 1, 2, and s refer to the upstream, downstream, and shock frame, respectively. The jump conditions for 90\arcdeg shocks are solutions to the following one-dimensional MHD conservation equations \citep{ken84}:
\begin{subequations} \label{mhd}
\begin{align}
[1+\sigma_1(1-Y)]
\gamma_{\rm 1s}\mu_1^*=\gamma_{\rm 2s}\mu_2^*~,\\
[1+\frac{\sigma_1}{2\beta_{\rm 1s}^2}
(1-Y^2)]
p_{\rm 1s}^*\mu_1^*+\frac{P_1^*}{p_{\rm 1s}^*}
=p_{\rm 2s}^*\mu_2^*+(\frac{n_1}{n_2})\frac{P_2^*}{p_{\rm 2s}^*}~,
\end{align}
\end{subequations}
where $\sigma_1=B_{\rm 1s}^2/(4\pi n_1\mu_1\gamma_{\rm 1s}^2)$ is the magnetization, with $B_{\rm 1s}$ as the transverse magnetic field, $\mu^*_1$ as the dimensionless specific enthalpy, and $Y$ as the ratio of shock frame magnetic fields or density \citep{ken84}, where $\mu^*_{\rm i} $ and $Y$ are defined by
\begin{subequations} \label{sjc}
\begin{align}
\mu_{\rm i}^*=1+(\frac{n_1}{n_{\rm i}})\frac{\tilde{\Gamma}_{\rm i}}{\tilde{\Gamma}_{\rm i}-1}
P_{\rm i}^* ~,\\
Y\equiv \frac{B_{\rm 2s}}{B_{\rm 1s}}=\frac{N_{\rm 2s}}{N_{\rm 1s}}=\frac{\gamma_{\rm 2s}p_{\rm 1s}^*}{\gamma_{\rm 1s}p_{\rm 2s}^*}
=\frac{\gamma_{\rm 2s}n_2}{\gamma_{\rm 1s}n_1}~.
\end{align}
\end{subequations}
Here $P^*$ denotes the dimensionless thermal pressure, $P^*=P/n_1mc^2$, and $\tilde{\Gamma}$ is the adiabatic index. In our simulations the upstream flow is considered a cold plasma, i.e., $P_1^*=0$, so that $\mu_1^*=1$. Solving Eq. (\ref{mhd}a) for $\mu_2^*$ and inserting the resulting expression into Eq. (\ref{mhd}b) leads to the following equations for an unmagnetized plasma \citep{zha05}:
\begin{subequations} \label{sgb}
\begin{align}
p_{\rm 1s}^{*2}=\frac{
(\gamma_{21}-1)
(\tilde{\Gamma}\gamma_{21}+1)^2
}
{
\tilde{\Gamma}(2-\tilde{\Gamma})
(\gamma_{21}-1)
+2
}~,\\
\gamma_{\rm 1s}^{2}=\frac {
(\gamma_{21}+1)
[
\tilde{\Gamma}(\gamma_{21}-1)+1]^2
}
{
\tilde{\Gamma}(2-\tilde{\Gamma})
(\gamma_{21}-1)
+2
}~.
\end{align}
\end{subequations}
The corresponding jump condition can then be determined by Eq. (\ref{sjc}b). Now we consider an unmagnetized jet with Lorentz factor $\gamma_{41}$ being decelerated by an ambient medium with density $n_1$. For such system,  indices 1, 2, 3, and 4 refer to the  unshocked ambient, shocked ambient, shocked jet, and unshocked jet, respectively. Relative to the CD, the above formalism can easily include the shocks associated with jet-ambient interactions. The speed of the CD can be determined by applying Eq. (\ref {mhd}b) at the interface to the ambient and jet plasmas \citep{nish09}. Hence, in the CD frame we have
\begin{equation} \label {interface}
\mu_1^*\gamma_{\rm 1CD}^2\beta_{\rm 1CD}^2+P_1^*
=\frac{n_4}{n_1}\mu_4^*\gamma_{\rm 4CD}^2\beta_{\rm 4CD}^2+P_4^* ~.
\end{equation}
The jet speed measured in the CD frame and the associated Lorentz factor are given by
\begin{subequations} \label {cds}
\begin{align}
\beta_{\rm 4CD}=\frac{\beta_{41}-\beta_{\rm 1CD}}{1-\beta_{41}\beta_{\rm 1CD}} ~,\\
\gamma_{\rm 4CD}=(1-\beta_{\rm 4CD}^2)^{-1/2} ~.
\end{align}
\end{subequations}
In our simulations, the jet and ambient medium are initially cold plasmas. Hence Eq. (\ref {interface}) reduces to 
\begin{equation} \label {mur}
\mu_{\rm r}\gamma_{\rm 1CD}^2\beta_{\rm 1CD}^2=\gamma_{\rm 4CD}^2\beta_{\rm 4CD}^2 ~,
\end{equation}
where $\mu_{\rm r}$ is the jet-to-ambient enthalpy ratio. Making use of Eqs. (\ref {cds}), after some algebra Eq. (\ref {mur}) gives
\begin{equation} \label {scd}
\beta_{\rm 1CD}=\frac{
\gamma_{41}\mu_{\rm r}^{1/2}
}{\gamma_{41}\mu_{\rm r}^{1/2}+1}\beta_{41} ~,
\end{equation}
which is the speed of ambient medium towards the CD, and thus the speed of the CD through the ambient medium. In the shocked ambient, once $\beta_{\rm 1CD}$ ($\gamma_{\rm 1CD}$) is determined from Eq. (\ref {scd}), by $\gamma_{21}\rightarrow\gamma_{\rm 1CD}$ Eqs. (\ref {sgb}) can be solved to find the speed of the shocked ambient (FS) in the ambient frame, $\beta_{\rm FS1}=-\beta_{\rm 1s}$. The jump condition for the FS is
\begin{equation} \label {fjc}
\frac{n_2}{n_1}=\frac{
\tilde{\Gamma}
\gamma_{\rm 1CD}+1
}{\tilde{\Gamma}-1} ~.
\end{equation}
In the shocked jet (RS), making use of Eq. (\ref{cds}a) allows the speed of CD in the jet frame to be calculated $\beta_{\rm CD4}=-\beta_{\rm 4CD}$. Under $1\rightarrow4$, $2\rightarrow3$ and $\gamma_{34}\rightarrow\gamma_{\rm 4CD}$ transformations in Eqs. (\ref {sgb}), we can calculate the speed of the RS in the jet frame, $\beta_{\rm RS4}=-\beta_{\rm 4s}$. In this region the jump condition becomes
\begin{equation} \label {rjc}
\frac{n_3}{n_4}=\frac{
\tilde{\Gamma}
\gamma_{\rm CD4}+1
}{\tilde{\Gamma}-1} ~.
\end{equation}
Finally, the speed of the RS in the ambient frame is given by 
\begin{equation} \label {rss}
\beta_{\rm RS1}=\frac{\beta_{41}-\beta_{\rm RS4}}{1-\beta_{41}\beta_{\rm RS4}} ~.
\end{equation}


\end{document}